\documentstyle[12pt]{article}

\begin{document}
\input{psfig.sty}
\begin{flushright}
\baselineskip=12pt
UPR-901-T \\
\end{flushright}

\begin{center}
\vglue 1.5cm
{\Large\bf Time-Like Extra Dimension and Cosmological Constant in Brane Models}
\vglue 2.0cm
{\Large Tianjun Li~\footnote{E-mail: tli@dept.physics.upenn.edu,
phone: (215) 898-7938, fax: (215) 898-2010.}}
\vglue 1cm
{ Department of Physics and Astronomy \\
University of Pennsylvania, Philadelphia, PA 19104-6396 \\  
U.  S.  A.}
\end{center}

\vglue 1.5cm
\begin{abstract}
We discuss the general models with one time-like 
extra dimension and parallel  
3-branes on the space-time $M^4 \times M^1$.
We also construct the general brane models or networks with $n$ space-like
and $m$ time-like extra dimensions and with constant bulk cosmological
constant on the space-time $M^4\times (M^1)^{n+m}$, and point out
that there exist two kinds of models with zero bulk cosmological constant:
for static solutions, we have to introduce time-like 
and space-like extra dimensions, and
for non-static solutions, we can obtain the models
with only space-like extra dimension(s). 
In addition, we give two simplest models explicitly,
 and comment on the 4-dimensional effective cosmological constant.
\\[1ex]
PACS: 11.25.Mj; 04.65.+e; 11.30.Pb; 12.60. Jv
\\[1ex]
Keywords: Cosmological Constant; Brane Models; Extra Dimensions

\end{abstract}

\vspace{0.5cm}
\begin{flushleft}
\baselineskip=12pt
December 2000\\
\end{flushleft}
\newpage
\setcounter{page}{1}
\pagestyle{plain}
\baselineskip=14pt

\section{Introduction}

Although the Standard Model is very succesful from experiments at
LEP and Tevatron, it has some unattractive features which
may imply the new physics. 
One of these problems is that the gauge forces and the gravitational force
are not unified. Another is
the gauge hierarchy problem. Previously, two solutions to the
gauge hierarchy problem have
been proposed: one is the idea of the technicolor and  compositeness
which lacks calculability, and the other is the idea
of supersymmetry.    

About two years ago, it was suggested that the large 
compactified extra dimensions may also be the solution to the
gauge hierarchy problem~\cite{AADD}, because a low ($4+n$)-dimensional
Planck scale ($M_X$) may result in the large 4-dimensional Planck scale 
($M_{pl}$) due
to the large physical volume ($V_p^n$)
 of extra dimensions: $M_{pl}^2 = M_X^{2+n} V_p^n$. In addition,
about one years ago, 
 Randall and Sundrum~\cite{LRRS} proposed another scenario
that the extra dimension is an orbifold,
and  the size of extra dimension is not large
but the 4-dimensional mass scale in the Standard Model is
suppressed by an exponential factor from 5-dimensional mass
scale because of the exponential warp factor. Furthermore,  they suggested that
the fifth dimension might be non-compact~\cite{LRRSN}, and there may exist only 
one
brane with positive tension at origin, but, there  exists the  gauge
hierarchy problem.  The remarkable aspect of
the second scenario is that it gives rise to a localized graviton field.
Recently, a lot of 5-dimensional models with 3-branes were built [4-5].
We constructed the 
general models with parallel 3-branes on the five-dimensional
space-time, and obtained that the 5-dimensional GUT scale on each brane
can be identified as the 5-dimensional Planck scale, but the
4-dimensional Planck scale is generated from the low 4-dimensional
GUT scale exponentially in our world. Furthermore, 
the models with codimension-1 brane(s) were constructed on 
the six-dimensional and higher dimensional space-time [6-8].  
  
In above model buildings,  all the models with warp factor in the
metric have negative bulk
 cosmological constant. However, in string theory, it is natural to
take the bulk cosmological constant to be zero since the tree-level 
vacuum energy in the generic critical closed string compactifications
(supersymmetric or not) vanishes. And the zero bulk cosmological
constant is natural in the scenario in which the bulk is supersymmetric
(though the brane need not be), or the quantum corrections to the
bulk are small enough to be neglect in a controlled expansion.
 Therefore, how to construct
the models with zero bulk cosmological constant is an important
question in the model buildings. One solution was proposed where a 
scalar $\phi$, whose bulk potential is vanished, 
is introduced~\cite{AHDKS}.
In this scenario, $\phi$ becomes singular at a finite distance along
the extra dimension and the warp factor in the metric vanishes at singularity.
The good aspect of this approach is that, the brane tension can be set
arbitrary. However, the $Z_2$ symmetric and 4-dimensional Poincare invariant
solution is unstable under the bulk quantum corrections, and any procedure
which regularizes the singurality will introduce the fine-tuning which
self-tuning is supposed to avoid~\cite{FLLN}. 
Moreover, the time dependent
solution might be a saddle point which is unstable to the expansion or 
contraction of 
brane world, and might not conserve the energy on the brane~\cite{PBJMCG}. 
By the way, the quantum solutions of brane worlds in the WKB
approximation were also discussed in~\cite{CNKO}.

We would like to explore how to construct the
brane models or networks with zero bulk cosmological constant and without
the bulk scalar $\phi$.  
After we construct the general models or networks with $n$ space-like
and $m$ time-like extra dimensions and with constant $(4+n+m)$-dimensional
cosmological constant on the space-time $M^4 \times (M^1)^{n+m}$, where 
$M^4$ is the four-dimensional Minkowski
space-time and $M^1$ is one-dimensional manifold with or without 
boundary, we find out that if we introduce the time-like extra dimension(s)
or time-dependent solutions, we can obtain the brane models or networks
with vanishing $(4+n+m)$-dimensional cosmological constant.

Time-like extra dimension is not a new 
subject~\cite{DIRAC}. Kaluza-Klein's six dimensional
model with two times and compact extra dimensions was also investigated
before~\cite{CIMENTO}.
And the experimental lower bounds on the posssible violation of unitarity
put a limit on the maximum radius of the internal time-like 
directions~\cite{DVALI}.
In addition, F-theory has one time-like extra dimension~\cite{VAFA}.
 Recently, the time-like extra dimension is  
considered~\cite{MCABK} in the brane world 
scenarios, too.
By the way, the two-time physics, suggested by I. Bars et al, has
a new sympletic gauge symmetry which indeed removes all the ghosts, establishes
the unitarity and causality, and play a role analogous to 
duality~\cite{IBARS}.

Here, we do not want to explore the solution to
 the problems arising from the time-like
extra dimension(s): unitarity and causality. We would like to open our mind in
the brane model buildings, and hopefully, those problems might be solved
in future study. 

In this paper, first, we discuss  the general 
models with parallel 3-branes and one time-like extra dimension on
the space-time $M^4 \times R^1$ in detail, and similarly, one
can construct the general models with one time-like extra dimension
on the space-time 
$M^4 \times R^1/Z_2$, $M^4 \times S^1$ and
$M^4 \times S^1/Z_2$. In fact, we can obtain the general 3-brane models
with one time-like extra dimension from the previous 3-brane models with
one space-like extra dimension in~\cite{LTJII}
by making the following transformaitons for the metric $g_{55}$, the sectional
bulk cosmological constant $\Lambda_i$ and the brane tension $V_i$
\begin{equation}
g_{55} \longrightarrow - g_{55} \ ~,~\,
\end{equation}
\begin{equation}
\Lambda_i \longrightarrow -\Lambda_i ~,~
V_i \longrightarrow - V_i ~.~\,
\end{equation}

Moreover, we
construct the general brane models or networks with $n$ space-like
and $m$ time-like extra dimensions, 
 and with constant $(4+n+m)$-dimensional cosmological constant on the space-time
 $M^4 \times (M^1)^{n+m}$. We also include the time ($t'$) 
 term in the conformal metric.
Furthermore, we point out that there exist 
two kinds of the models or networks with
warp factor in the metric and zero bulk cosmological constant:
one is static, the other is non-static. 
For static models, in order to obtain zero bulk cosmological
constant, we have to introduce space-like and time-like 
extra dimensions. And if we required that the sum
 of brane tensions
be zero, i. e., the 
4-dimensional effective cosmological constant is zero,
 in order to solve the gauge hierarchy problem,
we have to introduce at least one brane with negative 
tension, which can not be located at fixed point.
For non-static models, we can introduce
only space-like extra dimension(s).
However, if we required that
the brane which includes our world have positive tension,
 in order to solve the gauge hierarchy problem,
we have to introduce at least one brane with negative 
 tension, which can not be located at fixed point.
Moreover, in order to have vanishing effective 
4-dimensional cosmological constant, we have to 
fine-tune the parameters and adjust the set up of the 
branes in
the non-static models.
In addition, the non-static solutions might not be stable.
We also give two simplest models explicitly.

\section{ General 3-Brane Models with One Time-like Extra Dimension}
In this section, we would like to discuss the 
general models with one time-like extra dimension and parallel
3-branes on the space-time $M^4\times M^1$.
 Because those models are similar to the general models 
with one space-like extra 
dimension and parallel 3-branes in~\cite{LTJII}, we only discuss
the models whose fifth dimension is $R^1$ in detail, and then point out
the tiny difference between the models with one time-like extra dimension and
the models with one space-like extra dimension. 

Assuming we have $l+m+1$ parallel 
3-branes, and
their fifth coordinates are: $-\infty < \tau_{-l} < \tau_{-l+1} < ...< 
\tau_{-1} < \tau_0
< \tau_{1} < ... < \tau_{m-1} < \tau_{m} < +\infty$,  
we obtain the metric in each brane 
from the five-dimensional metric $g_{AB}$ where  
$A, B = \mu, \tau$ by restriction
\begin{equation}
\label{smmetric}
g_{\mu \nu}^{(i)} (x^{\mu}) \equiv g_{\mu \nu}(x^{\mu}, \tau=\tau_i) ~.~\,
\end{equation}
In this paper, we assume that $g_{ \mu 5} = 0 $ here. 

The classical action is given by

\begin{eqnarray}
S &=& S_{gravity} + S_B 
~,~\,
\end{eqnarray}
\begin{eqnarray}
S_{gravity} &=& 
\int d^4 x  ~d\tau~ \sqrt{g} \{- \Lambda (\tau) + 
{1\over 2} M_X^3 R \} 
~,~\,
\end{eqnarray}
\begin{eqnarray}
S_B &=& \sum_{i=-l}^{m} \int d^4 x \sqrt{-g^{(i)}} \{ {\cal L}_{i} 
-  V_{i} \} 
~,~\,
\end{eqnarray}
where $M_X$ is the 5-dimensional Planck scale,
$\Lambda (\tau)$ is the 5-dimensional cosmological constant, and 
$V_i$ where $i=-l, ..., m$ is the brane tension.
The $\Lambda(\tau)$ is defined as
\begin{eqnarray}
\Lambda (\tau) &=& \sum_{i=1}^m \Lambda_i \left(\theta (\tau-\tau_{i-1}) - 
\theta (\tau-\tau_i) \right)
+ \Lambda_{+\infty} \theta (\tau-\tau_m)
\nonumber\\&& +
\sum_{i=-l+1}^0 \Lambda_i \left(\theta (-\tau+\tau_i) - 
\theta (-\tau+\tau_{i-1}) \right)
+ \Lambda_{-\infty} \theta ( -\tau + \tau_{-l} ) 
~,~\,
\end{eqnarray}
where $\theta (x) = 1$ for $x \geq 0$ and $\theta (x) = 0$ for $x < 0$. So,
$\Lambda(\tau) $ is sectional constant.

The 5-dimensional Einstein equations for above action are
\begin{eqnarray} 
\sqrt{g} \left( R_{AB}-{1 \over 2 } g_{AB} R \right) &=& - \frac{1}{ M_X^3} 
[ \Lambda (\tau) \sqrt{g} ~g_{AB} +  
\nonumber\\&& \sum_{i=-l}^{m}
V_{i} \sqrt{-g^{(i)}} ~g_{\mu \nu}^{(i)} 
~\delta^\mu_A \delta^\nu_B ~\delta(\tau-\tau_i)  ] ~.~ \,
\end{eqnarray}
Assuming that there exists a solution that
 respects   4-dimensional 
Poincare invariance in the $x^{\mu}$-directions, one obtains
the 5-dimensional metric
\begin{eqnarray} 
ds^2 = e^{- 2 \sigma(\tau)} \eta_{\mu \nu} dx^{\mu} dx^{\nu}
 - d\tau^2 ~.~\, 
\end{eqnarray}
With this metric, the Einstein equations reduce to
\begin{eqnarray}
\sigma^{\prime 2} = {  {\Lambda (\tau) } \over { 6 M_X^3}} ~,~
 \sigma^{\prime \prime} =  - \sum_{i=-l}^{m}
{{V_{i}} \over\displaystyle {3 M_X^3 } } \delta (\tau-\tau_i)
~.~\, 
\end{eqnarray}

The general solution to  above differential equations is
\begin{equation}
\sigma (\tau) = \sum_{i=-l}^m k_i |\tau-\tau_i| + k_c \tau + c 
~,~\,
\end{equation}
where $k_c$ and c are constants, and $k_i \not= 0$ for $i=-l, ..., m$.
The relations between the $k_i$ and $V_i$, 
and the relations between the $k_i$  and $\Lambda_i$ are
\begin{equation}
V_i= - 6 k_i M_X^3
~,~
\Lambda_i= 6 M_X^3 (\sum_{j=i}^m k_j - \sum_{j=-l}^{i-1} k_j-k_c)^2
~,~\,
\end{equation}
\begin{equation}
\Lambda_{-\infty}= 6 M_X^3 (\sum_{j=-l}^m k_j-k_c)^2
~,~ \Lambda_{+\infty}= 6 M_X^3 (\sum_{j=-l}^m k_j+ k_c)^2
~.~\,
\end{equation}
Therefore, the five-dimensional cosmological constant is positive except the 
section(s) 
with zero bulk cosmological constant, then, for any point in 
$M^4 \times R^1$, 
 which is not belong to any brane and the section(s)
  with zero bulk cosmological constant, 
there is a neighborhood which is diffeomorphic to 
( or a slice of ) $dS_5$ space.
Moreover, the five-dimensional
cosmological constant and brane tensions should satisfy above equations. In 
order to obtain the finite
4-dimensional Planck scale, we obtain the 
constraints: $ \sum_{j=-l}^m k_j > | k_c |$. So, the
sum of brane tensions is negative.

The general bulk metric is
\begin{equation}
ds^2 = e^{-2 \sum_{i=-l}^m k_i |\tau-\tau_i| -2 k_c \tau - 2 c} 
\eta_{\mu \nu} dx^{\mu} dx^{\nu} -
d\tau^2
~.~ \,
\end{equation}

And the corresponding 4-dimensional Planck scale  is
\begin{equation}
M_{pl}^2 =  M_X^3 \left( T_{-\infty, -l} + T_{m, +\infty} + \sum_{i=-l}^{m-1} 
T_{i, i+1} 
\right) 
 ~,~ \,
\end{equation}
where 
\begin{equation}
T_{-\infty, -l} = {1 \over\displaystyle {2 \chi_{-\infty}}} e^{-2 \sigma 
(\tau_{-l})}
 ~,~ 
T_{m, +\infty} = {1 \over\displaystyle {2 \chi_{+\infty}}} e^{-2 \sigma 
(\tau_{m})}
 ~,~ \,
\end{equation}
if $\chi_{i, i+1} \neq 0$, then
\begin{equation}
T_{i, i+1} = {1 \over\displaystyle {2 \chi_{i, i+1}}} \left( e^{-2 \sigma 
(\tau_{i+1})}
 - e^{-2 \sigma (\tau_i)} \right)
 ~,~ \,
\end{equation}
and if $\chi_{i, i+1} =0$, then
\begin{equation}
T_{i, i+1} = (\tau_{i+1}- \tau_i) e^{-2  \sigma (\tau_i)}
 ~,~ \,
\end{equation}
where
\begin{equation}
\chi_{\pm \infty} = \sum_{j=-l}^m k_j \pm k_c ~,~ 
\chi_{i, i+1} =
\sum_{j=i+1}^m k_j - \sum_{j=-l}^i k_j - k_c
 ~.~ \,
\end{equation}

In addition, the four-dimensional GUT scale on the $i-th$ brane $M_{GUT}^{(i)}$ 
is
related to the five-dimensional GUT scale on the $i-th$ brane $M5_{GUT}^{(i)}$ 
by
\begin{equation}
M_{GUT}^{(i)} = M5_{GUT}^{(i)} e^{-\sigma (\tau_i)} 
 ~.~ \,
\end{equation}
In this paper, we assume that $ M5_{GUT}^{(i)} \equiv M_X $, 
for  $i =-l, ..., m$.

These models can be generalized to the models 
with $Z_2$ symmetry. Because of $Z_2$ symmetry, $k_c = 0$.
 There are two kinds of such models, one has odd number
of the branes, the other has even number of the branes. For the first one,
we just require that $k_{-i} = k_i$, $ \tau_{-i} = - \tau_i$,
 and $m=l$. For the second one,
we just require that $k_{-i} = k_i$, $\tau_{-i} = -\tau_i$,
 $m=l$, and $k_0=0$ (no number 0 brane). Furthermore, these models
 can also be generalized to the general models
  whose fifth dimension is $R^1/Z_2$,
 one just requires that $k_{-i} = k_i$, $ \tau_{-i} = - \tau_i$,
 $m=l$, then, introduces the equivalence classes: $ \tau \sim - \tau$
 and $i-th ~brane \sim  (-i)-th ~brane$. The only  trick point in 
 this case is that the brane tension $V_0$ is half of the original value, i. e., 
$V_0= 3 k_0 M_X^3$. And one may notice that, the
sum of brane tensions is negative and the sectional bulk cosmological
constant is non-negative.

Similarly, we can construct the general models with one time-like
extra dimension on the 
space-time $M^4 \times S^1$ and 
$M^4 \times S^1/Z_2$, as we have done in~\cite{LTJII}. In fact,
we can obtain the 3-brane models with one time-like extra dimension 
from previous 3-brane
models with one space-like extra dimension in~\cite{LTJII}
 by making the following transformaitons for the metric $g_{55}$, sectional bulk
cosmological constant $\Lambda_i$ and brane tension $V_i$
\begin{equation}
g_{55} \longrightarrow - g_{55} \ ~,~\,
\end{equation}
\begin{equation}
\Lambda_i \longrightarrow -\Lambda_i ~,~
V_i \longrightarrow - V_i ~.~\,
\end{equation} 

By the way, if
the fifth dimension is compact, the sum of brane tensions is zero.
And the gauge hierarchy problem can be solved in all above models, as we
have discussed in~\cite{LTJII}.

\section{Brane Models or Networks with Time-like 
and Space-Like Extra Dimensions}
In this section, we will construct the brane models or
networks with $n$ space-like
and $m$ time-like extra dimensions, and with constant 
($4+n+m$)-dimensional cosmological constant
on the space-time $M^4\times (M^1)^{n+m}$.
We also include the linear time ($t^{\prime}$) term in the 
conformal metric. The brane models or networks with zero 
bulk cosmological constant are 
special cases of the general brane models or networks.

Assume we have
$n$ space-like extra dimensions, and $m$ time-like extra dimensions,
the ordered coordinates for the whole
space-time are: $t^{\prime}$, $x^1$, $x^2$, $x^3$,$ y^1$,$ y^2$, ...,
$y^n$, $\tau^1$, $\tau^2$, ..., $\tau^m$
(Note that ($t^{\prime}$, $x^1$, $x^2$, $x^3$,$ y^1$,$ y^2$, ...,
$y^n$, $\tau^1$, $\tau^2$, ..., $\tau^m$) $\equiv$ ($0, 1, 2, ..., n+m+3$)). 
Along each extra dimension, we have
papallel ($2+n+m$)-branes, so, each brane is the hypersuface which is
determined by the algebraic equation $y^i = y^i_j$ or $\tau^i = \tau^i_j$.
And we assume that if $j < k$, then, $y^i_j < y^i_k$ or $\tau^i_j < \tau^i_k$. 
Because we require that the ($4+n+m$)-dimensional
 cosmological constant is a constant on the
whole space-time, along each extra dimension, the brane tensions of
parallel branes will have the same magnitudes except the brane tension of
the brane at boundary, which is half of that magnitude. 

In our notation, the ($4+n+m$)-dimensional metric is $g_{AB}$, and the
metric on each brane is obtained by restriction, for
example, the metric of the brane at $y^i=y^i_j$ is
\begin{eqnarray}
(g^{y^i}_j)_{\hat {A} \hat {B}} \equiv g_{\hat{A}\hat{B}}
(y^i = y^i_j)  ~,~\,
\end{eqnarray}
where $\hat{A},\hat{B} \neq 3+i$.
And the metric of the brane at $\tau^i=\tau^i_j$ is
\begin{eqnarray}
(g^{\tau^i}_j)_{\hat {A} \hat {B}} \equiv g_{\hat{A}\hat{B}}
(\tau^i = \tau^i_j) ~,~\,
\end{eqnarray}
where $\hat{A},\hat{B} \neq 3+n+i$.

The classical action is
\begin{eqnarray}
S &=& S_{gravity} + S_{BS} +S_{BT} ~,~\,
\end{eqnarray}
where
\begin{eqnarray}
 S_{gravity} &=& \int dt^{\prime} d^3x ~d^ny~d^m\tau~
 \sqrt{(-1)^{1+m} g} ~({1\over 2} M_X^{2+n+m} R -\Lambda)  ~,~\,
\end{eqnarray}
\begin{eqnarray}
 S_{BS} &=& -\int dt^{\prime} d^3x ~d^ny~d^m\tau~
 (\sum_{i=1}^{n} \sum_{j_i} \sqrt{(-1)^{1+m}g^{y^i}_{j_i}}
  ~V^{y^i}_{j_i}~ \delta(y^i -y^i_{j_i} )) ~,~\,
\end{eqnarray}
\begin{eqnarray}
 S_{BT} &=& -\int dt^{\prime} d^3x ~d^ny~d^m\tau~
 (\sum_{i=1}^{m} \sum_{j_i} \sqrt{(-1)^{m}g^{\tau^i}_{j_i}}
  ~V^{\tau^i}_{j_i}~ \delta(\tau^i -\tau^i_{j_i} )) ~,~\,
\end{eqnarray}
where $M_X$ is the $(4+n+m)$-dimensional Planck scale,
$\Lambda $ is the $(4+n+m)$-dimensional cosmological constant, and 
$V^{y^i}_{j_i}$ and $V^{\tau^i}_{j_i}$ are the brane tensions.

The Einstein equations arising from
 above action are
\begin{eqnarray}
G_{AB}&\equiv& R_{AB}-{1 \over 2 } g_{AB} R=  
{1\over {M_X^{2+n+m}}}~T_{AB}  ~,~\,
\end{eqnarray}
where
\begin{eqnarray}
T_{AB} = T^{gravity}_{AB} + T^{BS}_{AB} + T^{BT}_{AB} ~,~\,
\end{eqnarray}
where
\begin{eqnarray}
T^{gravity}_{AB} = - g_{AB} \Lambda ~,~\,
\end{eqnarray}
\begin{eqnarray}
T^{BS}_{AB} = - \sum_{i=1}^{n} \sum_{j_i} 
\sqrt{\frac{g^{y^i}_{j_i}}{g}}~
  V^{y^i}_{j_i} ~(g^{y^i}_{j_i})_{\hat A \hat B}~
\delta^{\hat A}_A ~\delta^{\hat B}_B~  \delta(y^i -y^i_{j_i}) ~,~\,
\end{eqnarray}
\begin{eqnarray}
T^{BT}_{AB} = - \sum_{i=1}^{m} \sum_{j_i} 
\sqrt{-\frac{g^{\tau^i}_{j_i}}{g}}~
  V^{\tau^i}_{j_i}~ (g^{\tau^i}_{j_i})_{\hat A \hat B}~
\delta^{\hat A}_A ~\delta^{\hat B}_B  ~\delta(\tau^i -\tau^i_{j_i}) ~.~\,
\end{eqnarray}

We assume the 
metric to be conformally flat and write it as
\begin{eqnarray}
ds_{4+n+m}^2 & = & \Omega^2 (-dt^{\prime 2} +\sum_{i=1}^3 dx^{i2}
+\sum_{i=1}^n dy^{i2} - \sum_{i=1}^m d\tau^{i2}) ~,~\, 
\end{eqnarray}
where $\Omega\equiv\Omega(t^{\prime}, y, \tau)$. The simplest 
way to proceed is to transform the metric to a conformally related 
metric, {\it i.e.} 
\begin{equation}
g_{AB} =\Omega^{2}~ \tilde g_{AB} ~. 
\end{equation}
The Einstein tensors in the two metrics are related by
\begin{eqnarray}
G_{AB} &=& \tilde G_{AB}+ (2+n+m)\left[  \tilde {\nabla}_{\! A} \ln \Omega 
 ~\tilde {\nabla}_{\! B} \ln \Omega - \tilde {\nabla}_{\! A} 
 \tilde {\nabla}_{\! B} \ln \Omega 
\right.\nonumber\\&&\left.
+ \tilde g_{AB}  \left(\tilde \nabla^2 \ln \Omega + 
\frac{1+n+m}{2} ( \tilde{\nabla} \ln \Omega)^2\right)\right]~,~\,
\end{eqnarray}
where the covariant derivatives  $\tilde{\nabla}$ are evaluated with respect to 
the metric $\tilde g$. Since the metric is conformally flat, 
 the covariant derivatives are identical to the ordinary derivatives and 
$\tilde G_{AB}=0$. So, the above equation can be recast to
\begin{eqnarray}
G_{AB} = (2+n+m)\left[ \Omega \tilde{\nabla}_{\! A} \tilde{\nabla}_{\! B} 
\Omega^{-1}
+  \tilde g_{AB} \left(- \Omega \tilde \nabla^2 \Omega^{-1} + 
\frac{3+n+m}{2} ~\Omega^2 ( \tilde{\nabla} \Omega^{-1})^2\right)\right]~.~\,
\end{eqnarray}

Using above form of Einstein tensor, the Einstein equations 
 can be written as 
\begin{eqnarray}
\frac{\partial^2}{\partial t^{\prime 2}} \Omega^{-1}& =& 0 ~,~ \,
\end{eqnarray}  
\begin{eqnarray}
\frac{\partial^2}{\partial y^{i2}} \Omega^{-1}& =&
\sum_{j_i} {1 \over\displaystyle {M_X^{2+n+m} (2+n+m)}}~V_{j_i}^{y^i}
 ~\delta(y^i-y^i_{j_i}) ~,~ \,
\end{eqnarray}
\begin{eqnarray}
\frac{\partial^2}{\partial \tau^{i2}} \Omega^{-1}& =&
-\sum_{j_i} {1 \over\displaystyle {M_X^{2+n+m} (2+n+m)}}~V_{j_i}^{\tau^i}
 ~\delta(\tau^i-\tau^i_{j_i}) ~,~ \,
\end{eqnarray}
\begin{eqnarray}
(\tilde \nabla \Omega^{-1})^2 &=& 
- { {2 \Lambda} \over\displaystyle {M_X^{2+n+m} (2+n+m)(3+n+m)}} ~.~\,
\end{eqnarray}

We can relate the fundamental scale $M_X$ to the four-dimensional 
Planck scale $M_{Pl}$ by
integrating over the extra dimensions 
\begin{eqnarray}
M_{Pl}^2 = M_X^{2+n+m} \int ~d^ny~d^m\tau~ \Omega^{2+n+m}~.
\end{eqnarray}
The four-dimensional ``Grand Unification Scale'' 
$(M^4_{GUT})_{(i_1, ..., i_n, j_1, ...,
j_m)}$  
at each brane junction is 
\begin{eqnarray}
(M^4_{GUT})_{(i_1, ..., i_n, j_1, ...,
j_m)} = M_X \Omega(y^1=y^1_{i_1},..., y^n=y^n_{i_n},
 \tau^1=\tau^1_{j_1}, ..., \tau^m=\tau^m_{j_m}) ~,~\, 
\end{eqnarray}
where we have assumed that the ($4+n+m$)-dimensional Planck
scale is equal to the ($4+n+m$)-dimensional GUT scale at
each brane junction.
Thus, it is possible to solve the gauge hierarchy
problem by choosing $\Omega$ appropriately.

Noticing that $\Omega^{-1}$ is a linear combination of the solutions to 
Eqs. (38-40), we obtain
\begin{equation}
\Omega^{-1} = h~ t^{\prime} + \sum_{i=1}^n \sigma^{Si}(y^i) 
+\sum_{j=1}^m \sigma^{Tj} (\tau^j) + c ~,~\,
\end{equation}
where  $\sigma^{Si} (y^i)$
and $\sigma^{Tj} (\tau^j)$ satisfy Eqs. (39) and (40), respectively.

Along each extra dimension, the brane tensions of
the parallel ($2+n+m$)-branes have the same magnitudes $|V^{y^i}|$ or 
$|V^{\tau^i}|$ except the brane tension of the brane at boundary,
which is half of that value, so, we define 
\begin{equation}
k^{y^i} = {1\over\displaystyle {2 (2+n+m) M_X^{2+n+m}}} |V^{y^i}| ~,~\,
\end{equation} 
and 
\begin{equation}
k^{\tau^j} = {1\over\displaystyle {2 (2+n+m) M_X^{2+n+m}}} |V^{\tau^j}| ~,~\,
\end{equation}
where $i=1, ..., n$ and $j=1, ..., m$.

Because in our discussion, the manifold of extra dimensions is the
product of $n+m$ one-dimensional manifold, each extra dimension can only be
$R^1$, $R^1/Z_2$, $S^1$ and $S^1/Z_2$. The functions of 
$\sigma^{Si} (y^i)$ and $\sigma^{Tj} (\tau_j)$ are 
the functions discussed in the second section in~\cite{LTJII}, so,
we just write down the results here.

For the space-like extra dimension, we use $y^i$ as an example:

(I) The 1-dimensional manifold for $y^i$ is $R^1$. 
Along $y^i$, we will 
have odd number of parallel ($2+n+m$)-branes. 
Assume we have $2 L +1 $ branes,
with coordinates on $y^i$: 
$-\infty <  y^i_{1} < ... < y^i_{ 2 L} < y^i_{ 2 L + 1} < +\infty$,
we obtain
\begin{equation}
\sigma^{Si} (y^i) = \sum_{j_i=1}^{2 L + 1} (-1)^{j_i+1} k^{y^i}
 |y^i-y^i_{j_i}|  
~,~\,
\end{equation}
and the brane tensions are
\begin{equation}
V^{y^i}_{j_i}= (-1)^{j_i+1} 2 (2+n+m) M_X^{2+n+m} k^{y^i} 
~.~\,
\end{equation}

(II) The 1-dimensional manifold for $y^i$ is $R^1/Z_2$. 
 Assume along $y^i$, we have $L +1 $ parallel ($2+n+m$)-branes,
with coordinates on $y^i$: 
$y^i_0=0 < y^i_1 <...<y^i_{L-1} < y^i_L < +\infty$, we have
\begin{equation}
\sigma^{Si} (y^i) = {1\over 2} (1+ (-1)^L) k^{y^i} |y^i| + 
\sum_{j_i=1}^{L} (-1)^{j_i+L} k^{y^i} |y^i-y^i_{j_i}|
~.~\,
\end{equation}
And the brane tensions for $j_i \not=0$ are
\begin{equation}
V^{y^i}_{j_i}= (-1)^{j_i+L} 2 (2+n+m) M_X^{2+n+m} k^{y^i} 
~,~\,
\end{equation}
and
\begin{equation}
V^{y^i}_{0}= (-1)^{L}  (2+n+m) M_X^{2+n+m} k^{y^i} 
~.~\,
\end{equation}

(III) The 1-dimensional manifold for $y^i$ is $S^1$. 
Along $y^i$, we will 
have even number of parallel ($2+n+m$)-branes. Assume we have $2 L $ branes,
with coordinates on $y^i$: 
$0= y^i_{1} < ... < y^i_{ 2 L-1} < y^i_{ 2 L } < 2 \pi \rho^i$,
where $\rho^i$ is the radius,
we obtain
\begin{equation}
\sigma^{Si} (y^i) = \pm \sum_{j_i=2}^{2 L } (-1)^{j_i+1} k^{y^i}
 |y^i-y^i_{j_i}|  
~,~\,
\end{equation}
and the brane tensions are
\begin{equation}
V^{y^i}_{j_i}= \pm (-1)^{j_i+1} 2 (2+n+m) M_X^{2+n+m} k^{y^i} 
~.~\,
\end{equation}
In addition, there is one constraint equation
\begin{equation}
\pm \sum_{j_i=2}^{2 L } (-1)^{j_i+1} y^i_{j_i} = - \pi \rho^i
~.~\,
\end{equation}

(IV) The 1-dimensional manifold for $y^i$ is $S^1/Z_2$. 
 Assume along $y^i$, we have $L + 1 $ parallel ($2+n+m$)-branes,
with coordinates on $y^i$: 
$0= y^i_{0} < ... < y^i_{L-1} < y^i_{ L } = \pi \rho^i$,
where $\rho^i$ is the radius,
we have
\begin{equation}
\sigma^{Si} (y^i) = \pm (\sum_{j_i=1}^{ L - 1 } (-1)^{j_i} k^{y^i}
 |y^i-y^i_{j_i}| + {1\over 2} (1+(-1)^{L+1}) k^{y^i} y^i)  
~.~\,
\end{equation}
And the brane tensions for $j_i \not= 0$ and $L$ are
\begin{equation}
V^{y^i}_{j_i}= \pm (-1)^{j_i} 2 (2+n+m) M_X^{2+n+m} k^{y^i} 
~,~\,
\end{equation}
and
\begin{equation}
V^{y^i}_{0}= \pm (2+n+m) M_X^{2+n+m} k^{y^i} 
~,~\,
\end{equation}
\begin{equation}
V^{y^i}_{L}= \pm (-1)^L (2+n+m) M_X^{2+n+m} k^{y^i} 
~.~\,
\end{equation}

Similarly, for the time-like extra dimension, 
we use $\tau^i$ as an example:

(I) The 1-dimensional manifold for $\tau^i$ is $R^1$. 
Along $\tau^i$, we will 
have odd number of parallel ($2+n+m$)-branes. Assume we have $2 L +1 $ branes,
with coordinates on $\tau^i$: 
$-\infty <  \tau^i_{1} < ... < \tau^i_{ 2 L} < \tau^i_{ 2 L + 1} < +\infty$,
we obtain
\begin{equation}
\sigma^{Ti} (\tau^i) = \sum_{j_i=1}^{2 L + 1} (-1)^{j_i+1} k^{\tau^i}
 |\tau^i-\tau^i_{j_i}|  
~,~\,
\end{equation}
and the brane tensions are
\begin{equation}
V^{\tau^i}_{j_i}= (-1)^{j_i} 2 (2+n+m) M_X^{2+n+m} k^{\tau^i} 
~.~\,
\end{equation}

(II) The 1-dimensional manifold for $\tau^i$ is $R^1/Z_2$. 
 Assume along $\tau^i$, we have $L +1 $ parallel ($2+n+m$)-branes,
with coordinates on $\tau^i$: 
$\tau^i_0=0 < \tau^i_1 <...<\tau^i_{L-1} < \tau^i_L < +\infty$, we have
\begin{equation}
\sigma^{Ti} (\tau^i) = {1\over 2} (1+ (-1)^L) k^{\tau^i} |\tau^i| + 
\sum_{j_i=1}^{L} (-1)^{j_i+L} k^{\tau^i} |\tau^i-\tau^i_{j_i}|
~.~\,
\end{equation}
And the brane tensions for $j_i \not=0$ are
\begin{equation}
V^{\tau^i}_{j_i}= (-1)^{j_i+L+1} 2 (2+n+m) M_X^{2+n+m} k^{\tau^i} 
~,~\,
\end{equation}
and
\begin{equation}
V^{\tau^i}_{0}= (-1)^{L+1}  (2+n+m) M_X^{2+n+m} k^{\tau^i} 
~.~\,
\end{equation}

(III) The 1-dimensional manifold for $\tau^i$ is $S^1$. 
Along $\tau^i$, we will 
have even number of parallel ($2+n+m$)-branes. Assume we have $2 L $ branes,
with coordinates on $\tau^i$: 
$0= \tau^i_{1} < ... < \tau^i_{ 2 L-1} < \tau^i_{ 2 L } < 2 \pi \rho^i$,
where $\rho^i$ is the radius,
we obtain
\begin{equation}
\sigma^{Ti} (\tau^i) = \pm \sum_{j_i=2}^{2 L } (-1)^{j_i+1} k^{\tau^i}
 |\tau^i-\tau^i_{j_i}|  
~,~\,
\end{equation}
and the brane tensions are
\begin{equation}
V^{\tau^i}_{j_i}= \pm (-1)^{j_i} 2 (2+n+m) M_X^{2+n+m} k^{\tau^i} 
~.~\,
\end{equation}
Moreover, there is one constraint equation
\begin{equation}
\pm \sum_{j_i=2}^{2 L } (-1)^{j_i+1} \tau^i_{j_i} = - \pi \rho^i
~.~\,
\end{equation}

(IV) The 1-dimensional manifold for $\tau^i$ is $S^1/Z_2$. 
 Assume along $\tau^i$, we have $L + 1 $ parallel ($2+n+m$)-branes,
with coordinates on $\tau^i$: 
$0= \tau^i_{0} < ... < \tau^i_{L-1} < \tau^i_{ L } = \pi \rho^i$,
where $\rho^i$ is the radius,
we have
\begin{equation}
\sigma^{Ti} (\tau^i) = \pm (\sum_{j_i=1}^{ L - 1 } (-1)^{j_i} k^{\tau^i}
 |\tau^i-\tau^i_{j_i}| + {1\over 2} (1+(-1)^{L+1}) k^{\tau^i} \tau^i)  
~.~\,
\end{equation}
And the brane tensions for $j_i \not= 0$ and $L$ are
\begin{equation}
V^{\tau^i}_{j_i}= \pm (-1)^{j_i+1} 2 (2+n+m) M_X^{2+n+m} k^{\tau^i} 
~,~\,
\end{equation}
and 
\begin{equation}
V^{\tau^i}_{0}= \pm (-1) (2+n+m) M_X^{2+n+m} k^{\tau^i} 
~,~\,
\end{equation}
\begin{equation}
V^{\tau^i}_{L}= \pm (-1)^{L+1} (2+n+m) M_X^{2+n+m} k^{\tau^i} 
~.~\,
\end{equation}

Furthermore, we obtain the $(4+n+m)$-dimensional
 cosmological constant
\begin{eqnarray}
\Lambda &=& -{1\over 2} (2+n+m) (3+n+m) M_X^{2+n+m}
\nonumber\\&&
(-h^2 + \sum_{i=1}^n (k^{y^i})^2 - \sum_{j=1}^m (k^{\tau^j})^2)
~.~\,
\end{eqnarray}

Requiring that 
$ -h^2 + \sum_{i=1}^n (k^{y^i})^2 - \sum_{j=1}^m (k^{\tau^j})^2 =0$,
we obtain the brane models or networks
 whose $(4+n+m)$-dimensional cosmological constant is zero. 
Therefore, there exist two kinds of models or networks with
warp factor in the metric and vanishing bulk cosmological constant: 
one is static ($h=0$), the other is non-static ($h\neq 0$).
For static models ($h=0$), we have to introduce space-like and time-like 
extra dimensions to obtain zero bulk cosmological
constant. And if one required that the sum
 of brane tensions
be zero,
i.e., the 4-dimensional effective cosmological constant is zero,
 in order to solve the gauge hierarchy problem,
one has to introduce at least one brane 
which has negative tension and can not be located at fixed point.
For non-static models ($h\neq 0$), we can introduce
only space-like extra dimension(s).
However, if one required that
the observable brane have positive tension,
 in order to solve the gauge hierarchy problem,
we have to introduce at least one brane which has negative tension and
 can not be located at fixed point.
Moreover, in order to have vanishing 4-dimensional effective
cosmological constant, we need to fine-tune the parameters
$h, k^{y^i}, k^{\tau^j}$ and adjust the set-up of the 
branes in the non-static models. 
In addition, the non-static solutions might not be stable.
 
By the way, using the $\sigma$ functions in the third section 
in~\cite{LTJII}, one can easily construct the general brane 
models or networks
with space-like and time-like extra dimensions whose
$(4+n+m)$-dimensional cosmological
constant is not constant on the whole space-time.

\section{Two Simplest Models} 
In this section, we will give two explicit simplest models 
with warp factor in the
metric and zero bulk cosmological constant.

(I)  We consider the static model with one space-like extra dimension $y^1$
and one time-like extra dimension $\tau^1$ on the space-time
$M^4 \times R^1 \times R^1$. For simplicity, we can write
$y$ and $\tau$ for the space-like extra dimension and time-like extra dimension
coordinates, respectively, i. e., $y \equiv y^1$,
$\tau \equiv \tau^1$. Because the solution is static, we have
$t^{\prime} \equiv t \equiv x^0$ and $h=0$. 
And we only consider two 4-branes, one brane with 
tension $V^y$ is the hypersuface determined by the equation $y=0$, the 
other brane with tension $V^{\tau}$ is the hypersuface determined 
by the equation $\tau = 0$. 
Zero bulk cosmological constant implies that $V^y= -V^{\tau}$. In short,
this is the simplest static model with warp factor in the metric
and zero bulk cosmological constant. The 
conformal metric is
\begin{eqnarray}
ds^2 & = & \Omega 
 (\eta_{\mu \nu} dx^{\mu} dx^{\nu}
+dy^2 - d\tau^2) ~,~\, 
\end{eqnarray}
where 
\begin{equation}
\Omega = {1\over\displaystyle {k|y| + k |\tau| +c}} ~,~\,
\end{equation}
where $c$ is a positive real number.
The brane tensions are
\begin{equation}
V^y = 8 ~M_X^4~ k
~,~ V^{\tau} = - 8~M_X^4~ k
~,~\,
\end{equation}
so, the 4-dimensional effective cosmological constant
is zero. The 4-dimensional Planck scale is
\begin{equation}
M_{pl}^2 = M_X^4 \int ~dy~ d\tau~ \Omega^4 
= {{2 M_X^4}\over\displaystyle {3 k^2 c^2}}~.~\,
\end{equation} 

Assuming the Standard Model lives at the
intersection of two branes, we obtain the 4-dimensional
GUT scale $M_{GUT}$ in our world
\begin{equation}
M_{GUT} ={{ M_X} \over c}  ~,~\,
\end{equation} 
therefore,
\begin{equation}
M_{pl} ={{\sqrt 2 M_X} \over\displaystyle {\sqrt 3 k}} M_{GUT}  ~.~\,
\end{equation}
We can not naturally solve the gauge hierarchy problem in this model. 
Of course, if we introduce more 4-brane(s), we can solve the gauge hierarchy
problem.

(II) The second model we consider is a non-static model with one space-like
extra dimension $y \equiv y^1$ on the space-time $M^4 \times R^1$. For 
simplicity, we 
just consider one 3-brane with tension $V$, which is the hypersuface determined 
by the equation $y=0$. Of course,
this is the simplest non-static model with warp factor in the metric
and zero bulk cosmological constant.
The conformal metric is
\begin{eqnarray}
ds^2 & = & \Omega 
 (-dt^{\prime 2} + \sum_{i=1}^3 dx^{i} dx^{i}
+dy^2 ) ~,~\, 
\end{eqnarray}
where 
\begin{equation}
\Omega = {1\over\displaystyle {h t^{\prime} +k|y| +c}} ~,~\,
\end{equation}
where $c$ is a real number. 
Requiring that the 5-dimensional cosmological constant is zero, we obtain 
$h=\pm k$.
The brane tension is
\begin{equation}
V = 6 ~M_X^3~ k
~,~\,
\end{equation}
so, the 4-dimensional effective cosmological constant is
non-zero. The 4-dimensional Planck scale is
\begin{equation}
M_{pl}^2 = M_X^3 \int ~dy~  \Omega^3 
= {{M_X^3}\over\displaystyle {k (h t^{\prime} + c)^2}}~.~\,
\end{equation} 

Assuming the Standard Model lives at 3-brane,
 we obtain the 4-dimensional GUT scale $M_{GUT}$ in our world
\begin{equation}
M_{GUT} = {{M_X} \over\displaystyle {h t^{\prime} + c}}  ~,~\,
\end{equation} 
therefore,
\begin{equation}
M_{pl}^2 ={{  M_X} \over\displaystyle {k}} M_{GUT}^2  ~.~\,
\end{equation}
We can not naturally solve the gauge hierarchy problem in this model. 
Of course, if one introduce more 3-brane(s), one
 can solve the gauge hierarchy problem. 

In addition, this model is unstable, in other words,
the universe either begins or ends in a singularity, depending on whether 
$h > 0$ or $h < 0$. But, if there exist higher order correction 
$(<t^{\prime}>)$ terms (non-linear $t^{\prime}$ terms) in the $\Omega^{-1}$,
for example $\epsilon ~t^{\prime 2}$, the solution might
be stable.

\section{Conclusion and Discussion}
 We construct the general 
models with parallel 3-branes and one time-like extra dimension on
the space-time $M^4 \times R^1$ in detail, and similarly, one
can discuss the general models with one time-like extra dimension
 on the space-time 
$M^4 \times R^1/Z_2$, $M^4 \times S^1$ and
$M^4 \times S^1/Z_2$. 
In addition, we
construct the general brane models or networks with $n$ space-like
and $m$ time-like extra dimensions, 
 and with constant 
$(4+n+m)$-dimensional cosmological constant on the space-time
 $M^4 \times (M^1)^{n+m}$. Time ($t'$) dependent term
 is also included in the conformal metric.
We point out that there exist two kinds of models or networks with
warp factor in the metric and zero $(4+n+m)$-dimensional cosmological constant: 
one is static, the other is non-static.
For static models, we have to introduce space-like and time-like 
extra dimensions to obtain vanishing bulk cosmological
constant. For non-static models, we can introduce
only space-like extra dimension(s).
We also give two simplest models explicitly.

Although taking zero bulk cosmological constant ($\Lambda =0$)
 is natural in the string theory
at tree-level or in the scenario where the bulk
is supersymmetric, one might expect the bulk quantum corrections to 
correct $\Lambda$ in a power series in the couplings, for example $g_s$. 
But, our solutions might still
be interesting if the bulk corrections to $\Lambda$ are very small, which
can happen for instance if the supersymmetry breaking is localized in
a small neighborhood of the branes, or if the supersymmetry breaking scale
in the bulk is small enough. Moreover, if all the gauge fields and matter fields
were confined to the branes, the quantum corrections of these fields to the
brane tensions might not affect the solutions with $\Lambda=0$, for example, in
the model I in section 4, if the quantum corrections to the two brane tensions
are equal, or in the model II in section 4, if the variation of $h$ 
 is equal to the quantum corrections to the brane tension. These results
are similar to those in the self-tuning models~\cite{AHDKS}. Of course,
we have fine-tuning in our solutions, but we do not have singularities.

\section*{Acknowledgments}
This research was supported in part by the U.S.~Department of Energy under
 Grant No.~DOE-EY-76-02-3071.


\begin{thebibliography}{99}
\itemsep 0.5mm
\bibitem{AADD} N. Arkani-Hamed, S. Dimopoulos and G. Dvali, Phys. Lett. 
B{\bf 429} (1998) 263, hep-ph/9803315; I. Antoniadis, 
N. Arkani-Hamed, S. Dimopoulos and G. Dvali, Phys. Lett. B{\bf 436}
 (1998) 257,
 hep-ph/9804398.
\bibitem{LRRS} L. Randall and R. Sundrum, hep-ph/9905221.
\bibitem{LRRSN} L. Randall and R. Sundrum, hep-ph/9906064.
\bibitem{JLLRLTJ} J. Lykken and L. Randall, hep-th/9908076;
 I. Oda, hep-th/9908104, hep-th/9909048;
 H. Hatanaka, M. Sakamoto, M. Tachibana, and K. Takenaga,
hep-th/9909076;
 T. Li, hep-th/9908174, hep-th/9911234.
\bibitem{LTJII} T. Li, hep-th/9912182.
\bibitem{HDDK} N. Arkani-Hamed, S. Dimopoulos, G. Dvali and N. Kaloper,
hep-th/9907209.
\bibitem{CCYS} C. Grojean, J. Cline and G. Servant, hep-ph/9909496;
 C. Csaki and Y. Shirman, hep-th/9908186;
 A. E. Nelson, hep-th/9909001; 
Z. Chacko and A. E. Nelson, hep-th/9912186;
 T. Torma, hep-th/0004021;
 S. Nam, hep-th/9911104;
 N. Kaloper, hep-th/9912125.
\bibitem{JJLTJ} J. Jiang, T. Li and D. Marfatia, hep-th/0007039.
\bibitem{AHDKS} N. Arkani-Hamed, S. Dimopoulos, N. Kaloper and R. Sundrum,
hep-th/0001197;
 S. Kachru, M. Schulz and E. Silverstein, hep-th/0001206, hep-th/0002121.
\bibitem{FLLN} S. Forste, Z. Lalak, S. Lavignac and H. P. Nilles,
hep-th/0002164,
hep-th/0006139.
\bibitem{PBJMCG} P. Binetruy, J. M. Cline and C. Grojean,
hep-th/0007029.
\bibitem{CNKO} L. Anchordoqui, C. Nunez, and K. Olsen, hep-th/0007064;
L. Anchordoqui, K. Olsen, hep-th/0008102.
\bibitem{DIRAC} P. A. M. Dirac, Ann. Math. {\bf 37} (1936) 429;
 H. A. Kastrup, Phys. Rev. {\bf 150} (1966) 1183;
 G. Mack and A. Salam, Ann. Phys. {\bf 53} (1969) 174.
\bibitem{CIMENTO} M. Pavisc, Nuovo Cimento {\bf B41} (1977) 397,
R. L. Ingraham, Nuovo Cimento {\bf B46} (1978) 1; {\bf B46} (1978) 16;
{\bf B46} (1978) 217; {\bf B46} (1978) 261; {\bf B47} (1978) 157;
{\bf B50} (1979) 233; {\bf B68} (1982) 203; {\bf B68} (1982) 218.
\bibitem{DVALI} G. Davali, G. Gabadadze and G. Senjanovic, {\it A
contribution to the Yu. A.  Golfand memorial volume}, Ed. M. A.
Shifman (World Scientific, 1999);
 F. J. Yndurain. Phys. Lett. {\bf B256} (1991) 15.
\bibitem{VAFA} C. Vafa, Nucl. Phys. {\bf B469} (1996) 403.
\bibitem{MCABK} M. Chaichian and A. B. Kobakhidze,
hep-th/0003269; M. Gogberashvili, hep-ph/0001109;
 M. Gogberashvili and P. Midodashvili, hep-ph/0005298.
\bibitem{IBARS} I. Bars, hep-th/0008164, and references therein;
 S. Vongehr, hep-th/9907034, and references therein.
 



\newpage








\end{thebibliography}
\end{document}